\documentclass[nobibnotes,twocolumn,aps,prl,superscriptaddress]{revtex4-1}
\usepackage{amsmath}
\usepackage{graphicx}
\usepackage[normalem]{ulem}
\usepackage{color}

\usepackage{natbib}
\begin{document}

\title{Fluid contact angle on solid surfaces: role of multiscale surface
roughness}
\author{F. Bottiglione}
\affiliation{Department of Mechanics, Mathematics and Management, Politecnico di Bari,
Italy}
\author{G. Carbone}
\affiliation{Department of Mechanics, Mathematics and Management, Politecnico di Bari
Italy\\
Department of Mechanical Engineering, Imperial College London, UK}
\author{B.N.J. Persson}
\affiliation{Peter Gr\"unberg Institut-1, FZ-J\"ulich, 52425 J\"ulich, Germany}
\affiliation{www.MultiscaleConsulting.com}

\begin{abstract}
We present a simple analytical model and an exact numerical study which
explain the role of roughness on different length scales for the fluid
contact angle on rough solid surfaces. We show that there is no simple
relation between the distribution of surface slopes and the fluid contact
angle. In particular, surfaces with the same distribution of slopes may
exhibit very different contact angles depending on the range of
length-scales over which the surfaces have roughness.
\end{abstract}

\maketitle



Controlling surface superhydrophobicity is of utmost importance in a
countless number of applications \cite{APPLICATIONS}. Examples are
anti-icing coatings \cite{anti-icing2,anti-icing3}, friction
reduction \cite{anti-friction1,anti-friction3}, antifogging
properties \cite{Gao-Mosquito}, antireflective coatings \cite{anti-reflective3}, solar cells \cite{solar-cells3}, chemical microreactors and
microfluidic microchips \cite{Hermingauschip}, self-cleaning paints
and optically transparent surfaces \cite{Blossey,Transp1,Transp5}. Water droplets on
superhydrophobic surfaces usually present very low contact angle hysteresis,
i.e. very low rolling and sliding friction values, which make them able to
move very easily and quickly on the surface, capturing and removing contamination particles, e.g, dust.
Surfaces with large contact angles, now referred to as 
exhibiting the ``lotus effect'', are found in many biological
systems, such as the Sacred Lotus leaves \cite{Bar} \cite{Barthlott}, water
striders \cite{waterstriders}, or mosquito eyes \cite{Gao-Mosquito}, where the presence of surface asperities cause the liquid
rest on the top of surface summits, with air entrapped between the drop
and the substrate. This type of ``fakir-carpet'' configuration is referred to
as the Cassie-Baxter state \cite{Cassie}. However, depending on the surface
chemical properties and its micro-geometry, a drop in a Cassie-Baxter state
may become unstable when the liquid pressure increases above a certain
threshold value\textit{\ }\cite{LafumaBIS,Carbone4,Carbone6,Carbone7,collaps-pressure, referee2-2,referee3-1,Persson2,Persson3}. 

The increase of liquid pressure may occur during drop impacts at high
velocities \cite{QuereRimb}. When this happens the
liquid undergoes a transition to the Wenzel state \cite{Wenzel}, which makes
the droplet rest in full contact with the substrate \cite{collaps-pressure,LafunaTransitio}. This transition to the Wenzel state may
be irreversible because of the energy barrier the drop should overcome to
come back to the Cassie-Baxter state \cite{LafumaBIS,LafunaTransitio,barrier,Carbone7}. However, the presence of
multi-scale or hierarchical micro-structures may favour the transition back
to the Cassie-Baxter state \cite{Reversible-switching,reversible4,referee1-1}, or
the stabilization of the `fakir-carpet' configuration \cite{Carbone7}, thus explaining why many biological systems present such a
multiscale geometry \cite{multiscale-real2,multiscale-real3}. 

Some studies \cite{Onda1,Onda2,Carbone1,Carbone2,Carbone3,Carbone5}, have shown that
many randomly rough surfaces possess super water-repellent properties with
contact angles up to 174${^{\circ }}$, which could explain why biological
systems use such hierarchical structures to enhance their hydrorepellent
properties\cite{fractal1}. Only few theoretical studies focus on this aspect
of the problem\cite{fractal1, Onda1, Onda2, fractal5,
fractal6}. In particular, it has been suggested \cite{Carbone1,Carbone2,Carbone3,Carbone5}\
that for randomly rough surface the critical parameter which stabilizes the
Wenzel or Cassie state is the so called Wenzel roughness parameter $r$.
Given Young's contact angle $\theta $ the Cassie state is stable when $r>-\left[ \cos \theta  \right] ^{-1}$, on the other hand when $r\leq -\left[ \cos \theta  \right] ^{-1}$ the low energy state
is the Wenzel state.
In this Letter we present exact numerical results, and the first analytical theory, which
show the fundamental role of roughness on many length scales 
for fractal-like surfaces, in generating large contact angles.

Consider a fluid droplet on a perfectly flat substrate. If the droplet is so
small that the influence of the gravity can be neglected, the droplet will
form a spherical cup with the contact angle $\theta $. In thermal
equilibrium the Young's equation is satisfied: 
\begin{equation*}
\gamma _{\mathrm{SV}}=\gamma _{\mathrm{SL}}+\gamma _{\mathrm{LV}}\mathrm{cos}\theta \eqno(1)
\end{equation*}

Consider now the same fluid droplet on a nominal flat surface with surface
roughness. If the wavelength $\lambda_0$ of the longest wavelength component
of the roughness is much smaller than the radius of the contact region, the
droplet will form a spherical cup with an (apparent) contact angle $\theta_0$
with the substrate, which may be larger or smaller than $\theta$ depending
on the situation. In this case the contact angle $\theta_0$ will again
satisfy the Young's equation, but with modified solid-liquid and solid-vapor
interfacial energies $\gamma^*_{\mathrm{SL}}$ and $\gamma^*_{\mathrm{SV}}$
(see also Ref. \cite{Persson2}): 
\begin{equation*}
\gamma^*_{\mathrm{SV}} = \gamma^*_{\mathrm{SL}}+\gamma_{\mathrm{LV}} \mathrm{cos}\theta_0 \eqno(2)
\end{equation*}

If the fluid makes complete contact at the droplet-substrate interface
(Wenzel model), then $\gamma^*_{\mathrm{SV}} =r \gamma_{\mathrm{SV}}$ and $\gamma^*_{\mathrm{SL}} =r\gamma_{\mathrm{SL}}$, where $r=A_{\mathrm{tot}}/A_0 $ is the ratio between the total surface area of the substrate
surface, and the substrate surface area projected on the horizontal $xy$-plane (also denoted nominal surface area $A_0$). Thus in the Wenzel model
assumption (2) takes the form 
\begin{equation*}
r \gamma_{\mathrm{SV}} = r \gamma_{\mathrm{SL}}+\gamma_{\mathrm{LV}} \mathrm{cos}\theta_0
\end{equation*}
and combining this with (1) gives 
\begin{equation*}
\mathrm{cos} \theta_0= r \mathrm{cos} \theta\eqno(3)
\end{equation*}
so that if $\theta > \pi/2$, $\theta_0>\theta$.

Note that the equations above for $\gamma^*_{\mathrm{SL}}$ and $\gamma^*_{\mathrm{SV}}$ depend on the assumption that the surface energies $\gamma_{\mathrm{SV}}$ and $\gamma_{\mathrm{SL}}$ are independent on the surface
slope which may be the case for amorphous solids but in general not for
crystalline solids.

Let $P(s)$ be the probability distribution of the absolute value of the
surface slopes $s=|\nabla h(\mathbf{x})|$. If the fluid make complete
contact with the substrate then\cite{Tosatti} 
\begin{equation*}
r=\int_0^\infty ds \ P(s) \left ( 1+s^2 \right )^{1/2} \eqno(4)
\end{equation*}
For a randomly rough surface it has been shown that 
\begin{equation*}
P(s) = -{\frac{2s }{s_0^2}} e^{-(s/s_0)^2} \eqno(5)
\end{equation*}
where $s_0$ is the root-mean-square slope. For a cosines profile $h(x,y)=h_0 
\mathrm{cos} (qx)$ (where $q=2 \pi /\lambda$, where $\lambda$ is the
wavelength) one obtain 
\begin{equation*}
P(s)= {\frac{2}{\pi}} \left ( 2 s_0^2-s^2\right )^{-1/2} \eqno(6)
\end{equation*}
for $s<s_0 \surd 2$ and $P(s)=0$ for $s>s_0\surd 2 $, where $s_0=h_0 q/\surd
2$ is the root-mean-square slope.

When the contact angle $\theta > \pi /2$, for surface roughness with large
enough slopes, a fluid droplet may be in a state where the vapor phase occur
in some regions at the nominal contact interface. This is denoted the Cassie
state and for this case it is much harder to determine the macroscopic
(apparent) contact angle $\theta_0$. In fact, it is likely that many
metastable Cassie states can form. In general, if several (metastable)
states occur, the state with the smallest contact angle will be the (stable)
state with the lowest free energy.

\begin{figure}[tbp]
\includegraphics[width=0.8\columnwidth]{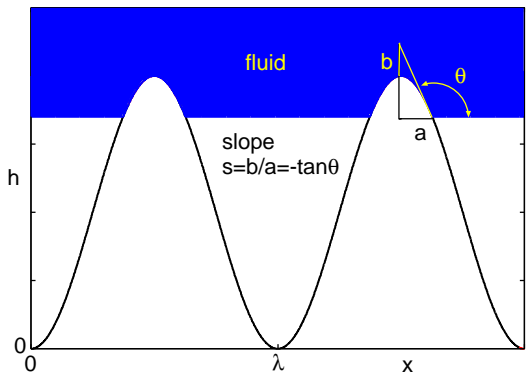}
\caption{Fluid in contact with a substrate with cosines corrugation. The
fluid occupy \textit{half} the region where the absolute value of the slope
is below $-\mathrm{tan}\protect\theta$. }
\label{sinus.ps}
\end{figure}

Let us first focus on the case where the fluid makes contact with a cosines
profile $h(x,y)=h_0 \mathrm{cos} (qx)$. In this case the contact will be as
indicated in Fig. \ref{sinus.ps}. If $-\mathrm{tan} \theta $ is larger than
the maximum slope $qh_0 = s_0\surd 2$, fluid will make complete contact with
the solid (Wenzel state). However, if $-\mathrm{tan} \theta < qh_0$ we have
the situation shown in Fig. \ref{sinus.ps}. In this case the fluid will make
contact with the solid in \textit{half} of the surface region where the
slope $s<-\mathrm{tan} \theta$ and we can write 
\begin{equation*}
\gamma^*_{\mathrm{SL}} =\int_0^{-\mathrm{tan} \theta} ds \ P(s) \left ({\frac{1}{2}} \left (\gamma_{\mathrm{SL}}+\gamma_{\mathrm{SV}} \right )\left
(1+s^2 \right )^{1/2}+ {\frac{1}{2}}\gamma_{\mathrm{LV}}\right )
\end{equation*}
\begin{equation*}
+\int_{-\mathrm{tan} \theta}^\infty ds \ P(s) \left (\gamma_{\mathrm{SV}}\left (1+s^2 \right )^{1/2}+\gamma_{\mathrm{LV}}\right ) \eqno(7)
\end{equation*}
and 
\begin{equation*}
\gamma^*_{\mathrm{SV}}=\gamma_{\mathrm{SV}} \int_0^\infty ds \ P(s) \left
(1+s^2 \right )^{1/2} \eqno(8)
\end{equation*}
Combining (2), (7) and (8) and using (1) gives 
\begin{equation*}
\mathrm{cos}\theta_0 = -1+ {\frac{1}{2}} \int_0^{-\mathrm{tan}\theta} ds \
P(s) \left (\left ( 1+s^2 \right )^{1/2} \mathrm{cos}\theta +1 \right ) \eqno(9)
\end{equation*}
If $\theta$ is close to $\pi$ we have $-\mathrm{tan}\theta \approx 0$ and we
can write (9) as 
\begin{equation*}
\mathrm{cos}\theta_0 = -1+ {\frac{1}{2}} \left ( \mathrm{cos}\theta +1
\right ) \int_0^{-\mathrm{tan}\theta} ds \ P(s) \eqno(10)
\end{equation*}
This equation was derived for a surface with a single cosines corrugation,
but should hold approximately also for a randomly rough surface with
roughness on a single length scale, i.e., a surface generated by the
superposition of cosines waves (or plane waves) with equal wavelength but
different propagation directions (in the $xy$-plane) and with different
(random) phases. Such a surface will have the distribution of slopes given
by (5). With $P(s)$ given by (5) from (10) we get 
\begin{equation*}
\mathrm{cos}\theta_0 = -1+ {\frac{1}{2}} \left ( \mathrm{cos}\theta +1
\right ) \left (1-\mathrm{exp} \left [\left ({\frac{\mathrm{tan}\theta }{s_0}}\right )^2 \right ]\right ) \eqno(11)
\end{equation*}
If we write $\theta_0 = \pi - \phi_0$ and $\theta = \pi - \phi$ with $\phi_0
<< 1 $ and $\phi << 1 $ we get from (11) 
\begin{equation*}
\phi_0 = {\frac{\phi^2 }{\surd 2 s_0}}\eqno(12)
\end{equation*}

\begin{figure}[tbp]
\includegraphics[width=0.8\columnwidth]{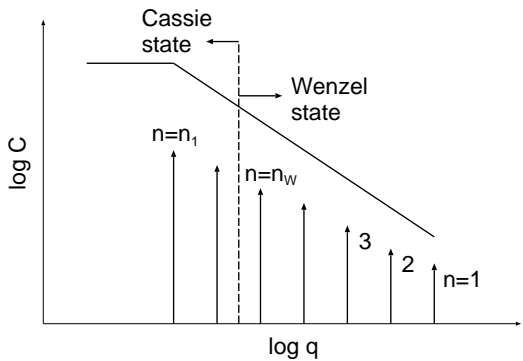}
\caption{The solid line shows the power spectrum as a function of the
wavevector (log-log scale) of a self-affine fractal surface with roughness
over several decades in length scale. The power spectrum given by the
vertical arrows (Dirac delta functions) correspond to a rough surface
composed of a discrete set of wavelength components. We can consider this as
a discretize version of the continuous power spectrum. We assume the
separation between two nearby delta functions to be of order 1 decade in
length scale. }
\label{qCpic.ps}
\end{figure}

\begin{figure}[tbp]
\includegraphics[width=0.6\columnwidth]{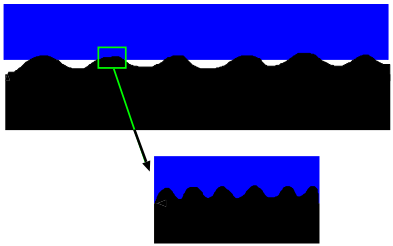}
\caption{The interface between a fluid droplet and a rough substrate. At
large length scales the liquid droplet will not make full contact with the
rough surface (Cassie state). However, when an apparent contact area is
magnified, it is observed that at shorter length scales the fluid is in
complete contact with the roughness profile. }
\label{CassieWenzel.ps}
\end{figure}

Consider a randomly rough surface with roughness on many length scales. Most
solids have surface roughness which is approximately self-affine fractal
over several decades in length scales\cite{Persson1}. A self affine fractal
surface has a power spectrum which depends on the wavevector $q$ as a power
law $C(q) \sim q^{-2(1+H)}$ where the Hurst exponent $0<H<1$. In addition,
most surfaces have a roll-off region for $q<q_{\mathrm{r}}$. The solid line
in Fig. \ref{qCpic.ps} shows the power spectrum of such a surface. Here we
consider instead a rough surface composed of a discrete set of wavelength
components with the power spectrum given by the vertical arrows (Dirac delta
functions) in Fig. \ref{qCpic.ps}. We can consider this as a discretize
version of the continuous power spectrum in Fig. \ref{qCpic.ps}. We assume
the separation between two nearby delta functions to be of order 1 decade in
length scale. This large separation in length scales allow us to integrate out, 
or eliminate the roughness which occur at length scales shorter than the 
length scale under consideration.

We will now consider how the contact angle $\theta_0$ changes as we
gradually add more roughness components to the surface profile. We will use
a Renormalization Group type of picture and study how the effective
interfacial energy $\gamma_{\mathrm{SL}}(n)$ (and $\gamma_{\mathrm{SV}}(n)$)
change as we include more and more of the surface roughness components,
i.e., we gradually increase $n$ from $n=1$ to the final value $n=n_1$ where
all the roughness components are included (see Fig. \ref{qCpic.ps}).
Consider first a surface where we only include the shortest wavelength
roughness components (associated the $n=1$ Dirac delta function in Fig. \ref{qCpic.ps}). Weassume that the slope of the surface is everywhere below $-\mathrm{tan}\theta$. In this case fluid will make complete contact with the
rough surface. If we now consider the system at a lower magnification the
surface appear smooth and flat. The fluid contact angle at this
magnification can be calculated from (3) with the modified (effective)
interfacial energy: 
\begin{equation*}
\gamma_{\mathrm{SL}}(n=1)=\gamma_{\mathrm{SL}} \times r_1
\end{equation*}
where $r_1=A(n=1)/A_0$ is the ratio between the surface area $A(n=1)$ and the nominal
surface area $A_0$. Using this effective interfacial energy, and a similar
expression for $\gamma_{\mathrm{SV}}(n=1)$, from (3) one obtain the fluid
droplet contact angle $\mathrm{cos}\theta(1)=r_1 \mathrm{cos}\theta$.
\begin{figure}[tbp]
\includegraphics[width=0.8\columnwidth]{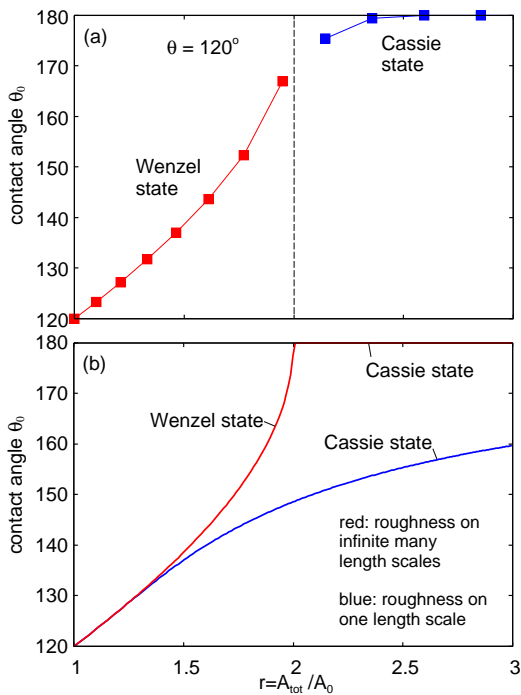}
\caption{The contact angle $\protect\theta_0 = \protect\theta(n)$ as a
function of the parameter $r=A_{\mathrm{tot}}/A_0$. (a) The red symbols are
the results obtained from the equation $\mathrm{cos}\protect\theta(n)=r_n 
\mathrm{cos}\protect\theta(n-1)$ with $n\le n_{\mathrm{W}}$ (in the present
case ($n\le n_{\mathrm{W}}=7$) and the blue symbols using $\protect\phi (n+1) = \protect\phi^2(n)/ (\surd 2 s_n)$ for $n>n_{\mathrm{W}}$. (b) The
limiting case where roughness occur on infinite many length scales (upper
curve) or on a single length scale (lower curve). }
\label{1Area.2theta0.theory.both.ps}
\end{figure}
Let us now add the next shortest wavelength roughness components,
represented by the Dirac delta function peek $n=2$ in Fig. \ref{qCpic.ps}.
We assume that the slope of the surface is everywhere below $-\mathrm{tan}\theta (1)$. In this case fluid will make complete contact with the rough
surface. If we now consider the system at a lower magnification the surface
appear smooth and flat. The fluid contact angle at this magnification can be
calculated from (3) with the modified (or effective) interfacial energy: 
\begin{equation*}
\gamma_{\mathrm{SL}}(n=2)=\gamma_{\mathrm{SL}}(1) \times r_2
\end{equation*}
where $r_2=A(n=2)/A(n=1)$ where $A(n=2)$ is the surface area
obtained with the $n=1$ and $n=2$ roughness components.
Using this effective interfacial energy, and a similar
expression for $\gamma_{\mathrm{SV}}(n=2)$, from (3) one obtain the fluid
droplet contact angle $\mathrm{cos}\theta(2)=r_2 \mathrm{cos}\theta(1)$.
More generally, we have 
\begin{equation*}
\mathrm{cos}\theta(n)=r_n \mathrm{cos}\theta(n-1)\eqno(13)
\end{equation*}
where $r_n = A(n)/A(n-1)$.
Eq. (13) is valid also for $n=1$ if we define $\theta(n=0)=\theta$ as the
contact angle on the flat perfectly smooth surface, and $A(n=0)=A_0$.
If the root-mean-square slope associated with the roughness in each of the
Dirac delta functions peeks is small enough, the procedure described above
can be continued until $\theta (n)$ is close to $\pi $ (or $180^{\circ }$)
and we conclude that for a surface with roughness on many length scales the
Wenzel state will prevail until the root-mean-square roughness is so high
that the Wenzel equation predict a contact angle close to $\pi $. We define $n=n_{\mathrm{W}}$ as the smallest $n$ for which $r_{n_{\mathrm{W}}+1}\mathrm{cos}\theta (n_{\mathrm{W}})<-1$, at which point there is no solution to the
Wenzel equation (Eq. (3)): $\mathrm{cos}\theta (n_{\mathrm{W}}+1)=r_{n_{\mathrm{W}}+1}\mathrm{cos}\theta (n_{\mathrm{W}})$

Consider now adding the $n=n_{\mathrm{W}}+1$ roughness component. Since the
maximum surface slope associated with the $n=n_{\mathrm{W}}+1$ roughness
component is larger than $-\mathrm{tan}\theta(n_{\mathrm{W}})$, the liquid
droplet will not make full contact with the rough surface. However, at
shorter length scales the fluid is in complete contact with the roughness
profile as indicated in Fig. \ref{CassieWenzel.ps}. To study the development
of the contact angle as we add more roughness we assume that (9) is valid,
which for $\theta=\pi-\phi \approx \pi$ takes the form (12) which we now
write as 
\begin{equation*}
\phi (n+1) = {\frac{\phi^2(n) }{\surd 2 s_n}}\eqno(14)
\end{equation*}
for $n\ge n_{\mathrm{W}}$, with $\phi (n_{\mathrm{W}})$ determined by the
contact angle in the Wenzel state which prevail when $n=n_{\mathrm{W}}$. In
(14) $s_n$ is the rms slope associated with the surface roughness contained
in the $n$'th Dirac delta function (see Fig. \ref{qCpic.ps}). Note that for
a randomly rough surface $s_n$ can be calculated from $r_n$.

We now present numerical results to illustrate the discussion above. Assume
that the fluid contact area on the flat smooth surface equals $\theta =
120^\circ$ as would be typical for Teflon. Assume that $r_1 = r_2 = .. = 1.1$. This correspond to $s_1=s_2=... \approx 0.46$. Note that the total surface
area after adding the first $n$ roughness wavelength components is $A_{\mathrm{tot}}/A_0 = r_1 \times r_2 \times ... \times r_n$. In Fig. \ref{1Area.2theta0.theory.both.ps}(a) we show the contact angle $\theta_0 =\theta(n)$ as a function of the parameter $r=A_{\mathrm{tot}}/A_0$.

It is instructive to consider two limiting cases, namely the case where
surface roughness occur over infinite decades in length scales (this limit
can of course not be realized in reality as it is meaningless to consider
roughness at length scales below the atomic dimension) and when roughness
occur on a single length scale. In the former case, we assume as above a
surface roughness power spectrum consisting of Dirac delta function peaks
separated by $\sim 1$ decade in lengthscale. We consider a surface with a
finite rms slope and we assume that each Dirac delta function contribute
equally to the rms slope. Hence since there are infinite many Dirac delta
functions each of them must have an infinitesimal weight corresponding to
roughness with infinitesimal amplitude. It is clear that in this case the
condition $r_{n_{\mathrm{W}}+1} \mathrm{cos}\theta(n_{\mathrm{W}}) < -1$
will be satisfied only when the contact angle $\theta(n_{\mathrm{W}})$
differ from $\pi$ by an infinitesimal amount, i.e., the Wenzel state will
prevail (minimum free energy state) until $r$ is so large that Eq. (3)
predict $\theta = \pi$ after which the Cassie state will prevail (with $\theta = \pi$).

For the second limiting case of roughness on a single length scale we use
(11) to estimate the contact angle $\theta_0$ for the Cassie state. In Fig.\ref{1Area.2theta0.theory.both.ps}(b) we show the contact angles for these
two limiting cases.

\begin{figure}[tbp]
\includegraphics[width=0.8\columnwidth]{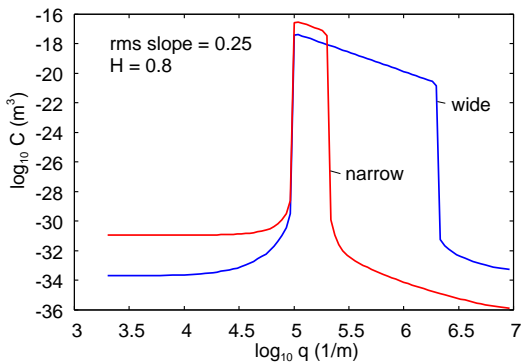}
\caption{The surface roughness power spectra $C(q)$ of surfaces with the rms
slope 0.25 and Hurst exponent $H=0.8$ but different large wavevector
cut-off. }
\label{1logq.2logC.wide.narrow.ps}
\end{figure}

\begin{figure}[tbp]
\includegraphics[width=0.8\columnwidth]{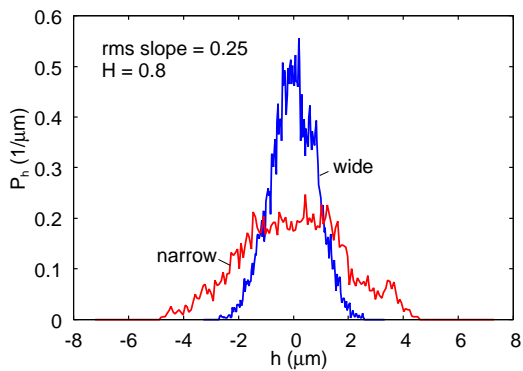}
\caption{The height distribution $P_h$ of surfaces with the rms slope 0.25
and Hurst exponent $H=0.8$ but different large wavevector cut-off. }
\label{1h.2Ph.wide.narrow.ps}
\end{figure}

We now present results for the contact angle as a function of the roughness
parameter $r$ based on an exact numerical treatment of an one-dimensional
(1D) model. Using standard procedures (see, e.g., Appendix D in Ref. \cite{Persson4}) we have generated randomly rough 1D surfaces and studied the
fluid contact angle using the method described in Ref \cite{Carbone2}. We
consider surfaces with roughness extending over what we denote as a
``narrow'' and a ``wide'' range of length scales. The roughness power
spectral density (PSD) $C(q)$ of the two types of surfaces are shown in Fig. \ref{1logq.2logC.wide.narrow.ps}. Both surfaces have the rms slope 0.25 and
Hurst exponent $H=0.8$, but different large wavevector cut-off. As a result,
the rms roughness is larger for the surface with the more narrow PSD. This
is illustrated in Fig. \ref{1h.2Ph.wide.narrow.ps} which shows the height
distribution $P_h$ of both surfaces.

We will show below that in spite of the larger rms roughness amplitude of
the surface with the more narrow PSD, the (apparent) contact angle is
largest on the surface with the widest PSD. This shows that the rms
roughness is irrelevant for the contact angle which instead is determined by
the rms slope and how different length scales contribute to the rms slope.

\begin{figure}[tbp]
\includegraphics[width=0.8\columnwidth]{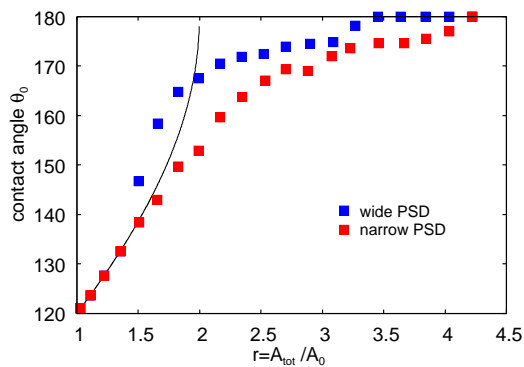}
\caption{The contact angle $\protect\theta_0$ as a function of the parameter 
$r=A_{\mathrm{tot}}/A_0$ obtained from numerical simulations. The solid line
is given by the Wenzel equation (3). }
\label{1Aratio.2ContAngle.exact1D.ps}
\end{figure}

Fig. \ref{1Aratio.2ContAngle.exact1D.ps} shows the contact angle $\theta_0$
as a function of the parameter $r=A_{\mathrm{tot}}/A_0$ obtained from
numerical simulations. Note that the surface with the more wide distribution
of roughness lengthscales has larger contact angle in spite of the fact that
both surfaces have the same rms slope. In general, several metastable
droplet states are possible, and which state occur depends on the
preparation procedure. We note, however, that the minimum free energy state
is the state with the lowest contact angle. Thus, the Cassie states observed
for $r < 2$ in Fig. \ref{1Aratio.2ContAngle.exact1D.ps} are metastable
states, and the ground state is in this case the Wenzel state. In the theory
presented above we assumed that the system was in the minimum free energy
state and no Cassie state could occur above the Wenzel line for $r<2$.

The result in Fig. \ref{1Aratio.2ContAngle.exact1D.ps} is consistent with
the theory prediction presented in Fig. \ref{1Area.2theta0.theory.both.ps}. In the
numerical simulations the contact angle in the Cassie state is smaller than
in the theory, but this reflects the fact that the roughness occur over a
wider range of length scales in the theory as compared to the numerical
model. This is indeed confirmed by numerical calculations presented by the
authors in \cite{Carbone2} (see Fig. 10 therein) where the surface roughness
was characterized by a much larger number of length scales covering about 3
decades.

To summarize, we have shown when the number of length scales of roughness
increases, the transition from the Wenzel state to the Cassie state approach
the threshold value $r=A_{\mathrm{tot}}/A_{0}=-1/\mathrm{cos}\theta $, and
the contact angle in the Cassie state approaches $180^{\circ }$.

\vskip 0.3cm

\bibliography{superhydrophobicity}
\end{document}